# VeriSFQ: A Semi-formal Verification Framework and Benchmark for Single Flux Quantum Technology

Alvin D. Wong, Kevin Su, Hang Sun, Arash Fayyazi, Massoud Pedram, Shahin Nazarian
Department of Electrical and Computer Engineering, University of Southern California, Los Angeles, CA
Email: {alvindwo, kevinsu, hangsun, fayyazi, pedram, shahin.nazarian}@usc.edu

**Abstract**

In this paper, we propose a semi-formal verification framework for single-flux quantum (SFQ) circuits called VeriSFQ, using the Universal Verification Methodology (UVM) standard. The considered SFQ technology is superconducting digital electronic devices that operate at cryogenic temperatures with active circuit elements called the Josephson junction, which operate at high switching speeds and low switching energy – allowing SFQ circuits to operate at frequencies over 300 gigahertz. Due to key differences between SFQ and CMOS logic, verification techniques for the former are not as advanced as the latter. Thus, it is crucial to develop efficient verification techniques as the complexity of SFQ circuits scales. The VeriSFQ framework focuses on verifying the key circuit and gate-level properties of SFQ logic: fanout, gate-level pipeline, path balancing, and input-to-output latency. The combinational circuits considered in analyzing the performance of VeriSFQ are: Kogge-Stone adders (KSA), array multipliers, integer dividers, and select ISCAS'85 combinational benchmark circuits. Methods of introducing bugs into SFQ circuit designs for verification detection were experimented with – including stuck-at faults, fanout errors, unbalanced paths, and functional bugs like incorrect logic gates. In addition, we propose an SFQ verification benchmark consisting of combinational SFQ circuits that exemplify SFQ logic properties and present the performance of the VeriSFQ framework on these benchmark circuits. The portability and reusability of the UVM standard allows the VeriSFQ framework to serve as a foundation for future SFQ semi-formal verification techniques.

**Keywords**

Single-flux Quantum (SFQ), UVM, ATPG, Benchmark

## 1. Introduction

The ongoing demand for energy-efficient and high-performance computing has driven the development of semiconductors since its early days, but with the conclusive end of Moore's Law and rising challenges to the physical scaling of CMOS devices [1], there is a significant need for new device technologies to continue beyond end-of-scaling CMOS technology. Developed in the late 1980s, a very promising family of "beyond-CMOS" devices is single flux quantum (SFQ) circuits [2]. Similar to how CMOS circuits are built with transistors (3-terminal devices) as their active elements, SFQ circuits are built using Josephson junctions (2-terminal devices) as their active components. When Josephson junctions are operated at cryogenic temperatures, these superconducting devices exhibit the Josephson effect – a phenomenon of a current called super-current that flows indefinitely long without any applied voltage. As a result, SFQ circuits benefit from Josephson junctions with high switching speeds on the order of picoseconds and low switching energy on the order of $10^{-19}$ joules, as demonstrated at 4.2 Kelvin [3]. The switching energy of Josephson junctions is two to three magnitudes lower than that of end-of-scaling CMOS devices [1]. Thus, SFQ circuits have the potential to achieve the computing demands for energy-efficient and high-performance circuits [4], [5].

Despite the advantages of SFQ devices over end-of-scaling CMOS technology, verification techniques for SFQ are not as advanced as those of CMOS due to key differences between SFQ and CMOS logic. Therefore, developing efficient and appropriate verification techniques for SFQ devices is necessary to reduce verification time and accelerate the process of finding bugs in SFQ circuit designs [6]. Previously developed verification techniques for SFQ circuits include formal verification with delay-based time frame modeling [7] and simulation-based verification using random high-speed testing [8], but to the best of our knowledge, none have focused on the third type of verification – semi-formal verification.

The main difference between CMOS and SFQ logic is the representation of binary information. CMOS logic uses voltage levels to represent "logic-1" (high) and "logic-0" (low), while SFQ logic stores information in a single quantum of superconducting flux $\Phi_0$, defined as

$$\Phi_0 = h/2e = 2.07 mV \times ps \qquad (1)$$

where $h$ is the Planck constant and $e$ is electron charge. SFQ typically appears as a voltage pulse with a peak amplitude of 2 to 4 millivolts and a duration of 1 to 2 picoseconds, thus the presence of an SFQ pulse is interpreted as "logic-1" and its absence "logic-0" [9].

The four key circuit and gate-level requirements of SFQ logic are related to: fanout, gate-level pipeline, path balancing, and input-to-output latency (or product of stage delay and circuit depth). Conventional SFQ logic requires that all circuit nets have only two pins, thus an SFQ gate can only drive one other node for a fanout of one. When a gate needs more than one fanout, a special SFQ gate called a splitter (see Fig. 1) is inserted at the output of that gate to create fanout for two gates. For additional fanouts, a binary tree of splitters is used where $n-1$ splitters are needed for $n$ fanouts.

Meanwhile, SFQ gate-level pipelining mandates that all SFQ gates, except splitters, should have a clock signal to transfer the stored quantum flux to their output and synchronize their operation. Prior work in clock distribution networks of SFQ circuits to achieve synchronization include [10], [11], and [12].

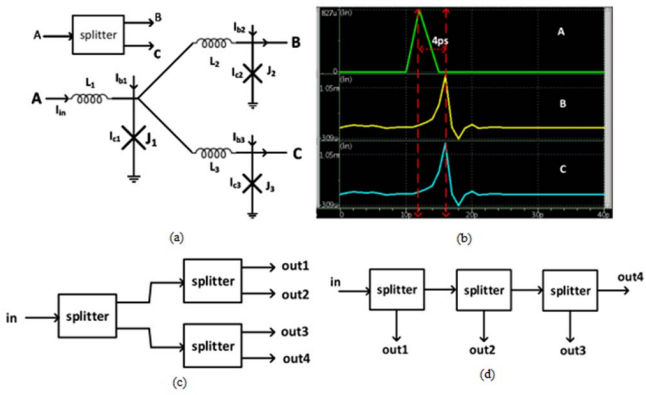

**Figure 1:** (a) SFQ Splitter Gate, (b) Waveform of Splitter Gate Operation, (c, d) Splitter Trees Providing Fanout of 4.

Thus, nearly all SFQ gates can be thought of as purely combinational gates followed by clocked D flip-flops (DFFs) and as such, every SFQ circuit must be completely gate-level pipelined. To ensure SFQ gate-level pipelining and maintain proper operation of SFQ gates, path balancing mandates that all inputs of an SFQ gate should have the same logic level[1]. If not, DFFs should be inserted into the path(s) of smaller depth to balance the network of inputs. Because of SFQ gate-level pipelining, the input-to-output latency of an SFQ circuit is defined as the product of the logical depth of the circuit and the clock cycle time, which is set by the worst-case single stage delay in the gate-level pipeline (gate delay plus any splitter delay plus interconnect delay).

In this paper, we present VeriSFQ, a semi-formal verification framework for SFQ circuits using the Universal Verification Methodology (UVM) standard. The UVM standard was chosen for our semi-formal verification because of its phasing structure for the lifecycle of a testbench, its monitoring and reporting mechanisms including functional coverage, its interfaces, its comprehensive base core libraries developed from the SystemVerilog hardware design and verification language, and the reusability and portability of UVM as a best-practice verification methodology [13]. Like in [14] for CMOS technology, using UVM has the potential to be embedded into a learning framework to increase the verification quality of future complex SFQ systems with potentially hard-to-detect design errors.

The VeriSFQ framework focuses on verifying the four key circuit and gate-level properties for SFQ logic. SFQ circuit fanout and path balancing are checked with pre-processing checkers before applying the design under verification (DUV) to the UVM environment, where gate-level pipelining and input-to-output latency of the DUV is maintained and monitored.

Lastly, our paper presents an SFQ verification benchmark comprised of combinational SFQ circuits that demonstrate the SFQ logic properties of fanout, gate-level pipeline, path balancing, and input-to-output latency – which future SFQ verification techniques should successfully detect.

The key contributions of this paper may be summarized as:

(i) Developing two checkers to verify the SFQ logic-specific requirements of fanout, gate-level pipeline, and path balancing for a combinational SFQ circuit.

(ii) Speeding up the verification process for SFQ circuits by utilizing UVM's phasing structure, its monitoring and reporting mechanisms for efficient bug detection, its modularity, and its portability, reusability, and scalability for future semi-formal SFQ verification techniques.

(iii) Simulating functional SFQ circuit design bugs for verification detection by modeling SFQ gates at "logic-0" with stuck-at zero faults.

(iv) Presenting an SFQ verification benchmark consisting of combinational SFQ circuits that exemplify the SFQ logic properties of fanout, gate-level pipeline, path balancing, and input-to-output latency.

The rest of the paper is organized as the following: Section 2 outlines the details of the VeriSFQ framework. The performance of the verification framework is discussed in Section 3. Section 4 presents an SFQ verification benchmark. Finally, the paper is concluded in Section 5.

## 2. VeriSFQ Framework

The system-level flowchart of VeriSFQ is depicted in Fig. 2. The SFQ circuits used to analyze the VeriSFQ framework are described by the SFQ cells and parameters detailed in [15] and [16]. First, our combinational SFQ circuit designs are structurally modeled at gate-level using the SFQ logic synthesis tool called SFQmap [17], [18]. Then, the structural SFQ circuit models are pre-processed to ensure proper fanout and path balancing before converting the structural model to its hardware description language (HDL) equivalent for UVM-compatibility. Also, the structural circuit models are fault modeled for single stuck-at faults (SSAFs) to simulate SFQ design bugs, and then analyzed by the academic Automatic Test Pattern Generation (ATPG) tool called ATALANTA [19] to generate minimal sets of test vectors that detect these faults. Finally, the circuit designs are semi-formally verified in UVM by applying ATPG-based test vectors or pseudo-random test vectors to their respective DUV and comparing the results against their "golden" SFQ circuit reference model.

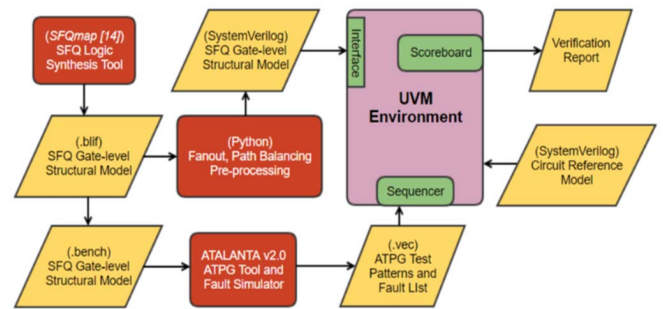

**Figure 2:** The Flowchart of Our SFQ Verification Framework, VeriSFQ.

---

[1]Logic level of gate *i* denotes the length of the longest path (in terms of the gate count) from any primary input (pi) of the network to this gate.

## 2.1. Adopting SSAF Modeling for SFQ Verification

Due to the nature of SFQ logic representing binary information with a pulse, SFQ gates functionally default to "logic-0" when not producing a pulse, i.e. in the absence of inputs that would logically produce a pulse and when waiting for the arrival of a pulse(s) on their input(s) that would logically produce a pulse. In these situations, an SFQ gate can be defined as inactive. As a functional bug, an inactive SFQ gate could indicate the failure to send or receive a pulse, i.e. an unconnected net in the design or an unconnected clock signal to the SFQ gate or a fan-in gate.

Thus, to simulate a functional bug, the behavior of an inactive SFQ gate can be modeled as a stuck-at zero fault on the interconnect of the gate output. Using ATPG, a minimal set of test vectors can be generated from this SSAF model and applied to the DUV to speed up verification time and functional coverage convergence, as opposed to using constrained random stimuli for functional verification. This modeling approach could accelerate the process of finding SFQ circuit design bugs as not all gates in a circuit are simultaneously active or change logical value when a new test vector is applied.

## 2.2. Pre-processing

Given a combinational SFQ circuit, the SFQ logic synthesis tool SFQmap [17], [18] generates its equivalent gate-level structural model. The VeriSFQ framework then extracts the circuit network of gates and wires from this model and analyzes the network to ensure the circuit has primary input fanout of one, SFQ gate fanout of one (adding splitters if needed for fanouts of more than one), and is entirely path balanced as a gate-level pipeline. To verify these properties, we utilize two checkers: an SFQ circuit fanout checker followed by an SFQ circuit path balancing checker; their pseudocode are detailed in Fig. 3 and 4, respectively.

The fanout checker ensures that once splitters are inserted to adjust the SFQ gate fanout for any gate output wire connected to two other gates, all primary inputs and gate output wires in the SFQ circuit are connected to one other gate. Meanwhile, the path balancing checker is a Depth First Search (DFS) that assumes the SFQ circuit is represented as a Directed Acyclic Graph (DAG). Thus, the DFS is constrained to iterate over just the primary inputs of the

```
Input: SFQ circuit's netlist
Output: Fanout error
1:  Extract lists of circuit's wire adjacency (WA) list as well as primary
    input (PI), gate (G), and primary output (PO)
2:  For each primary input pi in PI list
3:      If WA[pi]'s size is not exactly 1
4:          Return Fanout Error
5:  End For
6:  For each gate g in G list
7:      If gate g is a Splitter
8:          If WA[g]'s size is not exactly 1 or 2
9:              Return Fanout Error
10:     Else
11:         If WA[g]'s size is not exactly 1
12:             Return Fanout Error
13: End For
14: Return no Fanout Error
```

**Figure 3:** SFQ Circuit Fanout Checker Pseudocode.

```
Input: SFQ circuit's PI, G, and WA lists
Output: Path balancing error
1:  Connect all pi in PI to root r
2:  For each gate g in G                              // Initialization
3:      Set gate g's visited color as WHITE           // For unvisited
4:      Set gate g's level as 0
5:  End For
6:  Set max circuit depth mcd to 0
7:  Call SFQPathDepthCounter with path depth of 0 and root r
8:  If path depth list values are not all one value or do not match mcd
9:      Return Path Balancing Error
10: Else
11:     Return no Path Balancing Error
```
(a)

```
Function Name: SFQPathDepthCounter
Input: SFQ circuit's PO list, Current path depth pd, SFQ circuit node n
Output: Current path depth list
1:  If node n's visited color is BLACK                // For finished
2:      If pd + 1 equals node n's level
3:          Add mcd to n's path depth list
4:          Return n's path depth list
5:      Else
6:          Add mcd − |pd + 1 − node n's level| to n's path depth list
7:          // This shows the erroneous path depth count compared to mcd
8:          Return n's path depth list
9:  Else
10:     Set node n's visited color to GRAY            // For visited but not finished
11:     If node n is a Splitter
12:         Set node n's level to pd (note: splitters do not increase logic level or path depth)
13:         If n's adjacency list size is exactly 2
14:             Call SFQPathDepthCounter with node n's level and next node WA[n][0]          // WA is an adjacency list of lists
15:             Call SFQPathDepthCounter with node n's level and next node WA[n][1]
16:             Set node n's visited color to BLACK   // For finished
17:             Return WA[n][0]'s path depth list + WA[n][1]'s path depth list
18:         Else                                      // Adjacency size is 1
19:             Call SFQPathDepthCounter with node n's level and next node WA[n]              // WA[n] returns a 1-item list
20:             Set node n's visited color to BLACK   // For finished
21:             Return WA[n][0]'s path depth list
22:     Else
23:         Set node n's level to pd + 1
24:         If WA[n] is in PO
25:             If mcd < node n's level
26:                 Set mcd to node n's level
27:             Add node n's level to n's path depth list
28:             Return n's path depth list
29:         Else
30:             Call SFQPathDepthCounter with node n's level and next node WA[n]
31:             Set node n's visited color to BLACK   // For finished
32:             Return n's path depth list
```
(b)

**Figure 4:** Pseudocode for (a) SFQ Circuit Path Balancing Checker, (b) SFQ Circuit Path Depth Counter.

circuit. The recursive DFS visits have been modified to handle splitters as splitters do not increase logic level or path depth but present two separate paths to recurse on.

For SFQ circuits with proper fanout, the fanout checker has a worst-case runtime complexity of $O(|PI| + |G|)$, where $|PI|$ is the number of primary inputs and $|G|$ is the number of SFQ gates in the circuit (including splitters). Meanwhile, the path balancing checker and path depth counter have a worst-case runtime complexity of $O(|PI| + |G| + |W|)$, where $|W|$

is the number of wires in the SFQ circuit. Both the fanout checker and path balancing checker's worst-case runtime complexity reduce to O(|G|) because each wire connects to up to two gates and the number of primary inputs is typically less than the number of gates in a circuit.

Upon passing both checkers, the structural model is converted into the SystemVerilog HDL for UVM-compatibility as the DUV. Otherwise, the SFQ circuit design is flagged for the corresponding fanout or path balancing error and rejected from undergoing verification in UVM, thus saving verification time.

### 2.3. SFQ Logic Simulation

Defined to approach real-life behavior of an SFQ logic signal, an interface was used to simulate how SFQ logic represents binary information [12]. The interface consisted of the logic value, a sent flag, a pulse width parameter, a send task to send a pulse of set width and set the flag, a receive task to read the incoming data on its positive edge, and a receiving function to obtain the data and reset the flag. Using the signal interface, SFQ gates were defined as clocked modules – which include a buffer, splitter, inverter, and 2-input AND, OR, and XOR gates. Upon receiving data at its inputs, these SFQ gates generate corresponding outputs on the clock's next positive edge.

### 2.4. UVM Testbench Architecture

Following the best-practice verification methodology of the UVM standard, our testbench components are built from the UVM library's component classes. Thus, our testbench architecture is portable and scalable to other SFQ circuit designs.

#### 2.4.1. Data Items (Transactions) and DUV Interface

Our testbench defines its stimuli (transactions in UVM) as data packets containing the primary inputs and outputs of the SFQ circuit (DUV), which are interfaced using the SFQ signal interface.

#### 2.4.2. Sequencer and Sequences

Depending on the type of sequence desired for design bug detection, our sequencer applies either transactions of constrained random test vectors or the minimal set of ATPG test vectors to the DUV. To reduce simulation time and accelerate functional coverage convergence, constrained randomization over the $2^n$ possible number of inputs, i.e. pseudo-randomization using random seeds and application of the $2^n$ possible number of inputs, was used instead of direct (incremental) testing, i.e. incrementally applying 0 to the $2^n - 1$ possible number of inputs. Generally, in verification, direct (incremental) testing increases functional coverage at a steady rate, but constrained randomization has shown faster convergence and steeper initial rate of convergence at the onset of verification.

#### 2.4.3. Driver and Monitor

Our driver reads the transactions generated by the sequencer and upon reading a "logic-1", the send task of the corresponding SFQ bit is called from the DUV interface to generate the desired voltage pulse. Fork-join statements group each data packet to ensure DUV inputs are generated in parallel and DUV outputs are driven in the same clock. At every clock, our monitor samples the DUV inputs and outputs for the scoreboard to collect coverage information, generating a "logic-1" on the corresponding bit upon detection of a voltage pulse by reading for the positive edge of the pulse.

#### 2.4.4. Scoreboard and Reference Model

Our scoreboard compares the actual output from the DUV with the expected output from the reference model. Our reference models for combinational SFQ circuits are pipelined with depth equal to each circuit's input-to-output latency. Thus, transactions are applied to the DUV in consecutive clocks, and outputs for those transactions are read and compared after the circuit's input-to-output latency time has passed. Hence, the VeriSFQ framework has a runtime complexity of O($l \times (t + 1)$), where $l$ is the circuit's input-to-output latency and $t$ is the number of applied transactions.

## 3. Experimental Results

In this section, the efficacy of VeriSFQ is investigated. First, the simulation setup and considered combinational SFQ circuits are described. Then, a case study comparing the use of SSAF modeling and incorrect logic gates as functionally erroneous designs is discussed by examining the verification detectability variance of VeriSFQ and comparing functional coverage convergence when using either ATPG-based test vectors or pseudo-random test vectors to detect these simulated design bugs. Finally, the performance of the VeriSFQ framework in detecting these simulated SFQ circuit design bugs is presented.

### 3.1. Simulation Setup and Combinational SFQ Circuits

All UVM verification framework simulations were performed in Mentor Graphics' QuestaSim 64-bit version 10.4c. The computer system used for the simulations utilized an Intel Core i7-6650U (duo-core) CPU with nominal clock frequency of 2.2 gigahertz and 8 gigabytes of RAM.

The considered combinational SFQ circuits include Kogge-Stone adders (KSA), array multipliers, integer dividers, and ISCAS'85 combinational benchmark circuits c17 and c6288 [20] and [21]; their essential statistics are shown in Table 1.

### 3.2. VeriSFQ Detection Results

Two models of introducing functional bugs into SFQ circuit designs were experimented with for verification detection: stuck-at zero faults to model inactive SFQ gates and incorrect logic gate usage. Each method was verified separately in sample runs as a new DUV and compared with the corresponding "golden" SFQ circuit reference model. The VeriSFQ framework can run two verification engines based on the type of test vector applied: pseudo-random test vectors (Engine 1) and ATPG-based test vectors (Engine 2).

#### 3.2.1. Verification Detectability Variance Case Study of SSAF Modeling and Incorrect Logic Gate Usage

A case study was conducted to observe the verification detectability variance of VeriSFQ when using single stuck-at fault modeling and incorrect logic gates to simulate functionally erroneous SFQ circuit designs and to also compare the application of ATPG-based test vectors and pseudo-random test vectors to the framework. Fig. 5 shows the cumulative distribution function (CDF) of the time to first detected error for the 32-bit KSA using both verification engines on the two functionally erroneous design models, while Fig. 6 shows the functional coverage convergence of both verification engines for the 32-bit KSA. The 32-bit KSA was chosen as a representative SFQ circuit because of its relative complexity and size combined with its lower input-to-output latency.

Table 1: Combinational SFQ Circuit Characteristics.

| SFQ Circuit Name | # of Primary Inputs | # of Primary Outputs | # of DFFs | # of Splitters | # of INV/ AND/ OR/ XOR Gates | Input -to- Output Latency |
|---|---|---|---|---|---|---|
| KSA4 | 9 | 5 | 27 | 28 | 32 | 6 |
| KSA8 | 17 | 9 | 80 | 71 | 79 | 8 |
| KSA16 | 33 | 17 | 220 | 178 | 194 | 10 |
| KSA32 | 65 | 33 | 580 | 437 | 469 | 12 |
| ArrMult4 | 8 | 8 | 126 | 64 | 64 | 16 |
| ArrMult8 | 16 | 16 | 734 | 320 | 320 | 40 |
| ArrMult16 | 32 | 32 | 3390 | 1408 | 1408 | 88 |
| IntDiv4 | 8 | 8 | 309 | 99 | 123 | 27 |
| IntDiv8 | 16 | 16 | 2293 | 370 | 460 | 93 |
| c17 | 5 | 2 | 6 | 3 | 7 | 4 |
| c6288 | 32 | 32 | 3051 | 1431 | 1707 | 73 |

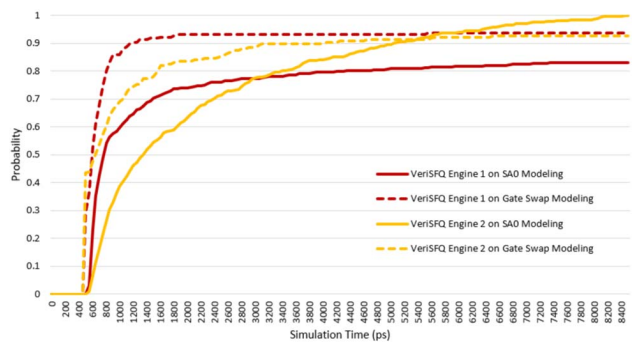

**Figure 5:** Verification Detectability CDF of Time to 1st Error Detected (KSA32) for Various Verification Engines of VeriSFQ.

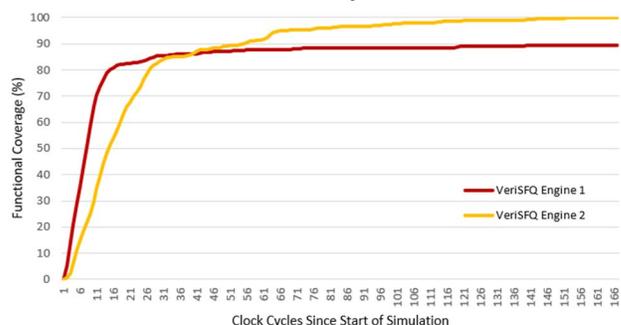

**Figure 6:** Functional Coverage Convergence (KSA32) for Various Verification Engines of VeriSFQ.

The CDFs in Fig. 5 show that the use of incorrect logic gates in SFQ circuits as functionally erroneous designs is much more easily detected than the use of SSAF modeling for inactive SFQ gates (dashed vs. solid CDFs). Interestingly, the CDFs in Fig. 5 also show that in verification time, bugs are more likely to be discovered sooner by the pseudo-random test vectors of Engine 1 as opposed to the ATPG-based test vectors of Engine 2. Yet, the ATPG-based test vectors of Engine 2 were able to achieve full convergence in a shorter period of time as opposed to the pseudo-random test vectors of Engine 1. Therefore, framework performance was investigated using verification detection sample runs conducted with SSAFs modeling functionally erroneous SFQ circuit designs, where Engine 2's ATPG-based test vectors outperforms Engine 1 because of its ability to find the bug. Meanwhile in Fig. 6, Engine 2 yields fuller but slower initial functional convergence as opposed to Engine 1. To define functional coverage of the VeriSFQ framework, covergroups containing all nets of an SFQ circuit were defined for toggle coverage using UVM. Selection of the sizeable and complex 32-bit KSA as a representative SFQ circuit reveals that pseudo-randomization alone saturates at a certain coverage and nearly stops converging due to the large possible number of inputs. Thus, the respective CDF in Fig. 5 covers a very long verification time that is beyond what is depicted because the sample runs eventually time out before the functional bugs are found.

Future work in the pseudo-randomization of inputs would include leveraging UVM's constrained randomization capabilities in VeriSFQ, as the verification framework is capable of applying different weights to input bits and using assertions – immediate and concurrent for combinational and sequential SFQ circuits. In addition, a more comprehensive functional coverage can be defined by leveraging the many types of UVM covergroups and coverpoints.

### 3.2.2. VeriSFQ Performance on SSAF-Modeled Design Bugs

The fault statistics produced by the academic ATPG tool ATALANTA [19] for the considered combinational SFQ circuits when fault modeled for SSAFs are shown in Table 2.

Table 2: Combinational SFQ Circuit Fault Characteristics.

| SFQ Circuit Name | Collapsed Faults | Redundant Faults | Aborted Faults | Fault Coverage (%) | # of ATPG Test Vectors |
|---|---|---|---|---|---|
| KSA4 | 146 | 0 | 0 | 100 | 18 |
| KSA8 | 350 | 0 | 0 | 100 | 32 |
| KSA16 | 842 | 0 | 0 | 100 | 66 |
| KSA32 | 1994 | 0 | 0 | 100 | 160 |
| ArrMult4 | 312 | 1 | 0 | 99.68 | 15 |
| ArrMult8 | 1520 | 1 | 0 | 99.93 | 33 |
| ArrMult16 | 6624 | 1 | 14 | 99.77 | 75 |
| IntDiv4 | 452 | 10 | 0 | 97.79 | 18 |
| IntDiv8 | 1666 | 3 | 0 | 99.82 | 44 |
| c17 | 22 | 0 | 0 | 100 | 7 |
| c6288 | 6744 | 2 | 95 | 98.56 | 90 |

To inject design bugs for verification detection in each sample run, an SSAF was inserted randomly on a net in the SFQ circuit using random seeding before being applied as a new DUV to the VeriSFQ framework. The SSAFs had a 99% chance of being a stuck-at zero fault to simulate the design bug of an inactive SFQ gate. For each SFQ circuit, CPU time to first functional error detected along with its standard deviation and relative error of the sample verification runs are shown in Table 3. Relative error tends to decrease as the combinational SFQ circuits grow in complexity, indicating less detectability variance with growing circuit complexity.

## 4. VeriSFQ Benchmark

In this section, a benchmark is presented that consists of combinational SFQ circuits suitable for evaluating the verification framework and exemplifying SFQ logic properties. Table 4 presents the essential statistics of the golden SFQ circuits, along with the statistics of the same circuits that were modified to be functionally erroneous in their SFQ logic properties of either fanout or gate-level pipeline and path balancing. Thus, the benchmark of modified SFQ circuits differs from the golden circuits in only DFF and splitter counts, as no functional gates were altered. In Table 4, the modified circuits termed fanout had fanout bugs inserted, whereas the ones designated unbalanced had a single path's input-to-output latency changed from the overall golden circuit. The insertion of these bugs involved the removal of a splitter to increase a gate fanout or the addition or removal of a DFF to shorten the input-to-output latency. The performance of the VeriSFQ framework on these SFQ logic design bugs is optimized by its SFQ logic property pre-processing checkers of fanout and path balancing.

Future work would include verification of more complex combinational SFQ circuits like an ALU, which would require more efficient and advanced reference models, and investigating the power of a machine learning-based framework (like in [14]) for future complex SFQ systems with potentially hard-to-detect design errors. We would also like to apply the VeriSFQ framework to verify the functionality of post-place and route netlists, e.g. after using [22], [23] as a tool for placement and routing.

**Table 3:** VeriSFQ Performance on SSAF-Modeled Bugs.

| SFQ Circuit Name | Pre-processing CPU Time (s) | Average Verification CPU Time to 1st Error (s) | Verification CPU Time Standard Deviation (±s) | Verification CPU Time Relative Error (%) |
|---|---|---|---|---|
| KSA4 | 0.035 | 0.082 | 0.016 | 19.57 |
| KSA8 | 0.040 | 0.101 | 0.029 | 28.53 |
| KSA16 | 0.064 | 0.096 | 0.028 | 28.71 |
| KSA32 | 0.114 | 0.190 | 0.087 | 46.00 |
| ArrMult4 | 0.044 | 0.087 | 0.024 | 27.16 |
| ArrMult8 | 0.096 | 0.225 | 0.044 | 19.75 |
| ArrMult16 | 0.359 | 1.118 | 0.129 | 11.51 |
| IntDiv4 | 0.063 | 0.105 | 0.019 | 17.83 |
| IntDiv8 | 0.171 | 0.625 | 0.023 | 3.74 |
| c17 | 0.034 | 0.079 | 0.028 | 35.52 |
| c6288 | 0.372 | 0.981 | 0.044 | 4.52 |

**Table 4:** VeriSFQ Benchmark Circuit Characteristics and Verification Framework Performance.

| SFQ Circuit Name | # of DFFs | # of Splitters | Input-to-Output Latency | Pre-processing CPU Time (s) |
|---|---|---|---|---|
| KSA4_golden | 27 | 28 | 6 | 0.035 |
| KSA4_fanout | 27 | 27 | 5*, 6 | 0.037 |
| KSA4_unbalanced | 26 | 28 | 5*, 6 | 0.034 |
| ArrMult4_golden | 126 | 64 | 16 | 0.044 |
| ArrMult4_fanout | 126 | 63 | 15*, 16 | 0.042 |
| ArrMult4_unbalanced | 125 | 64 | 15*, 16 | 0.049 |
| IntDiv4_golden | 309 | 99 | 27 | 0.063 |
| IntDiv4_fanout | 309 | 98 | 27 | 0.050 |
| IntDiv4_unbalanced | 308 | 99 | 26*, 27 | 0.053 |
| c17_golden | 6 | 3 | 4 | 0.034 |
| c17_fanout | 6 | 2 | 4 | 0.028 |
| c17_unbalanced | 5 | 3 | 3*, 4 | 0.036 |

\* The varied input-to-output latency produced by unbalancing the circuits.

## 5. Conclusion

In this work, a semi-formal verification framework for SFQ circuits named VeriSFQ is proposed. The SFQ logic-focused framework incorporates checkers for fanout, gate-level pipeline, and path balancing requirements. It also includes modular, scalable, and reusable test modules following the UVM standard. The performance of the VeriSFQ framework was investigated with a series of combinational SFQ circuits: adders, multipliers, dividers, and select ISCAS'85 benchmark circuits. Functionally erroneous designs were simulated using SSAFs to model inactive SFQ gates and incorrect logic gates for detection in verification. An SFQ verification benchmark exemplifying SFQ logic properties, including golden circuits and the same circuits containing fanout and path balancing errors, was also presented for future SFQ verification frameworks to successfully verify.

## 6. Acknowledgment

The research is based upon work supported by the Office of the Director of National Intelligence (ODNI), Intelligence Advanced Research Projects Activity (IARPA), via the U.S. Army Research Office grant W911NF-17-1-0120. The views and conclusions contained herein are those of the authors and should not be interpreted as necessarily representing the official policies or endorsements, either expressed or implied, of the ODNI, IARPA, or the U.S. Government. The U.S. Government is authorized to reproduce and distribute reprints for Governmental purposes notwithstanding any copyright notation herein.